# A numerical study on temperature destratification induced by bubble plumes in idealized reservoirs

**Yiran Li[a,b,d], Dongming Liu[a,b,c,*]**

[a] State Key Laboratory of Hydraulic Engineering Intelligent Construction and Operation, Tianjin University, Tianjin 300072, China

[b] School of Civil Engineering, Tianjin University, Tianjin 300072, China

[c] State Key Laboratory of Hydraulics and Mountain River Engineering, Sichuan University, Chengdu 610065, China

[d] China Railway First Survey and Design Institute Group Co., LTD, Xi'an 710000, China

[*]**Author to whom correspondence should be addressed**: hpeliudm@tju.edu.cn

## ABSTRACT

An in-house numerical model has been developed to study the temperature destratification induced by bubble plumes in reservoirs. The mean flow of the mixed fluid phase is solved using the Reynolds-Averaged Navier-Stokes Equations, while the bubble phase is calculated by employing the advection-diffusion equation of air concentration. The change of water temperature is tracked by solving the energy equation. The two-equation turbulence model, considering the effect of bubble buoyancy and the change of temperature, is adopted to model the turbulent dissipation in two-phase fluid. To validate the accuracy of the numerical model, the numerical results are compared with available experimental data. In addition, the destratification of temperature-layered water in a tank by bubble plumes is simulated by the numerical model, which is also validated by experimental data and the numerical results of another model. This three-dimensional model can reflect various physical quantities in the entire computed filed, such as thermal structure, gas concentration distribution, velocity distribution and turbulence intensity. Furthermore, by utilizing the validated model, a series of factors on reservoir destratification, including the aeration rate and the aeration location are analyzed and discussed in this study. A non-dimensional number $N_t$ is introduced to determine the optimal aeration rate of the fasted destratification. The extent to which bubble plumes can affect the destratification is also simulated and discussed. Meanwhile, the mechanism of destratification in a large water area is described in detail. Suggestions on how to apply bubble plumes to improve the water quality of a reservoir are given.

# 1 Introduction

With the rapid development of the social economy and accelerating urbanization in China and other developing countries, the water shortage and water quality deterioration are increasingly serious. In 2012, 63% of the world's inland water bodies were eutrophic, accounting for 31% of all water bodies. In 2019, among the 107 lakes and reservoirs monitored in China, 5.6% were classified as middle-eutrophic (the trophic level index (TLI):$60 \leq$ TLI $(\sum) \leq 70$), 23.4% as light-eutrophic ($50 \leq$ TLI $(\sum) \leq 60$), 61.7% as mesotrophic, and 9.3% as oligotrophic (Yan et al. 2021). Water eutrophication has become an increasingly serious problem worldwide (Moal et al. 2019). It has been widely acknowledged that water-temperature stratification is probably a primary cause of water pollution and eutrophication. Stratification can lead to cold water pollution, dissolved oxygen depletion, elevated concentrations of iron (Fe) and manganese (Mn) and can provide favorable conditions for algae growth (Chaaya and Miller 2022). Therefore, artificial destratification is key to controlling water quality.

A number of methods have been employed for artificial destratification, including mechanical stirrers, water pumps and bubble plumes. Bubble plumes, produced by releasing air through the diffuser located near the bottom of a reservoir, have been suggested as an important method to improve water quality (Koberg and Ford, 1965; Hughes et al., 1975). As air bubbles rise towards the free surface, the cold hypolimnetic water is carried upward with them. The carried water will be detrained from the bubbles either at the surface or at the level of neutral buoyancy where the density of air-water mixture is approximately equal to the ambient density. The detrained water can intrude, propagate into, then mix with the ambient water, ultimately reducing the thermal stratification. The efficiency of destratification basically depends on the aeration condition and the stratification structure. Although the detrainment of cold water can occur more than once as bubbles rise to the surface, it is preferred to design an aeration system that ensures the detrainment occurs at the free surface. If water area is too large for a single aeration to mix the stratified water completely, multiple devices should be considered. However, maximum efficiency of destratification by bubble plumes is usually achieved when individual bubble plumes do not interact with each other. As a result, it is important to determine the condition under which individual bubble plumes can reach the optimal mixing efficiency.

Many researchers have devoted their efforts to studying the properties of bubble plumes in water. For example, McDougall (1978) carried out experiments in a Perspex tank to study bubble plumes and the results showed that when bubbles keep rising, the water can leave the plume and spread out horizontally. Kranenburg (1978) performed experiments in a cylindrical tank and demonstrated that the bubble plumes can entrain the ambient water, forming an intrusion at the free surface, where the cold water mixes with epilimnetic water and there will be an interlayer of intermediate density between epilimnion and hypolimnion. Goossens (1979) also conducted a series of experiments at different scales and his experimental results had a good agreement with the findings by Kranenburg (1978). Milgram and van Housten (1982) performed experiments in their laboratory tank and the profiles of flow speed, momentum flux and gas concentration were measured and fitted with Gaussian curves. Godon and Milgram (1987) carried out experiments in a cubical tank and found that the time required for temperature stabilization was primarily controlled by the aeration rate, the plume length and the size of tank base, but not by tank height. Asaeda and Imberger (1988) performed experiments to study the destratification in both linearly stratified and a two-layered ambient fluids and they defined three types of intrusions depending on the aeration rate with respect to the stratification strength. Zic et al. (1989) carried out experiment to analyze the flow patterns and entrainment rates. Lemckert and Imberger (1993) conducted field and laboratory experiments to investigate the mixing induced by bubble plumes in water columns of arbitrary stable stratification, developing formulas for the radial extent of the plunge point and the net detrainment rate may for use in air-diffuser design producers. Seol et al. (2009) studied flow properties of a bubble plume in density-stratified conditions using planar laser-induced fluorescence (PLIF) flow visualization, particularly measuring the peel and intrusion heights for the first detrainment event above the diffuser. Through these experimental studies, we can have initial understanding about the mechanism of destratification by bubble plumes.

With the development of computer technology, many scholars have adopted and developed numerical models to study the destratification induced by bubble plumes. McDougall (1978) first developed an integral model of the bubble plume in stratified environments. The single-plume model was applied by Schladow (1992) to study the energy transfer of the destratification. Asaeda and Imberger (1993) modified the double-plume model, considering the inner plume as a rising part and the outer plume as a surrounding downdraft. Imteaz and Asaeda (2000) presented a composite model including both the mixing of lake water by air bubble plume and the lake phytoplankton module to predict and assess lake eutrophication. Singleton et al. (2007) improved the linear bubble plume model, which is derived from the circular model of Wüest et al. (1992) and modified by McGinnis et al. (2001), using full-scale diffuser data collected from Spring Hollow Reservoir (SHR) in the United States. Helfer et al. (2011) applied the one-dimensional process-based model DYRESM (Imberger and Patterson, 1981) to simulate lake mixing dynamics under artificial destratification and predict water temperatures and evaporation rates. Norman and Revankar (2011) used one-dimensional buoyant jet and two-phase turbulent jet-plume models to analyze the mixing dynamics and thermal response for shallow submergence of the source in large pools. Chu and Prosperetti (2017) used a standard horizontally integrated, quasi-one-dimensional model for bubble plumes in a stratified environment to study the aspects of the physics such as source parameters, scaling intrusion height and neutral height. The models mentioned above can simulate the change in water temperature during aeration, but they cannot reflect such physical quantities as gas concentration distribution, velocity distribution and turbulence intensity to provide detailed instantaneous plume structures for studying the fundamental physics. Given that three-dimensional (3D) computational fluid dynamics (CFD) models can reflect not only complex environmental conditions but also the distribution of physical parameters, the demand on the use of a 3D model has been growing rapidly. Fraga and Stoesser (2016) employed a large-eddy simulation based Eulerian-Lagrangian (EL) model to quantify the impact of bubble size, diffuser diameter, and gas flow rate on integral properties of bubble plumes. Zhou et al. (2020) also presented an EL large-eddy simulation of bubble plumes in linearly stratified environments, which treats bubbles as individual Lagrangian particles. However, those EL models may cause a high computational cost when the total number of bubbles in the computational domain is very big. Lee et al. (2013) evaluated the performance of a 3D hydrodynamic model, ELCOM, and simulated heat fluxes and stratification under different hydrological years of 2008 and 2009 using an extensive field dataset.

Birt et al. (2021) adopted AEM3D, a 3D model of hydrodynamics and ecology, coupling ELCOM and CAEDYM to consider various hydro-environments and capture related physical and biogeochemical processes. However, ELCOM is not a full 3D model because the hydrostatic pressure assumption is adopted in this model. When hydrodynamic pressure becomes significant during aeration process, the results may be of poor accuracy. Chen et al. (2017) used a semi-implicit 3D hydrodynamic model, Si3d, coupled with a newly developed water-jet model and an existing linear bubble-plume model to analyze a set of water quality management systems. The coupled models could provide data for understanding the mixing effects on water temperature, but not gas concentration in the whole field. Bravo et al. (2007) used the commercial software FLUENT, and constructed a two-fluid dispersed model to simulate the dynamics of bubble plumes. Xiao (2021) used the Eulerian-Eulerian (EE) LES bubble plume model developed in Yang *et al.* (2016) to study characteristics and material transport of a bubble-driven plume in stably stratified water with uniform crossflow. Fabregat et al. (2017) employed a simplified, EE model for bubble plumes, which was presented in Fabregat et al. (2015) and further investigated in stratified environments by Fabregat et al. (2016). The model adopted the Boussinesq assumption and neglected the momentum of gas phase. Compared with the two-fluid dispersed and EE models, it is more suitable for mixture model to implement large-scale and long-term simulations because of the lower computational complexity and cost. Therefore, recently the mixture model has been recommended widely to study two-phase fluid.

The objective of this study is to develop a mixture model for two-phase fluid, combining with the energy and turbulence models, which will be employed to simulate the effect of destratification by bubble plumes in a reservoir. First, based on the 3D Numerical Wave TANK (NEWTANK), which was developed by Liu and Lin (2008), the mixture model and the $k$-$\varepsilon$ model are adopted to calculate the mixed flow velocity and gas concentration distribution. The energy equation is utilized to track the change in temperature. Second, the mixture model and the energy model are validated by experiment data from Zarrati (1994) and Zic (1990). Furthermore, different situations are calculated to analyze the impact factors on destratification, including the aeration rate, the aeration location, and the extent to which bubble plumes can affect. Finally, suggestions on application of bubble plumes are given. The present study is, to the best of our knowledge, the first to use a 3D hydrodynamic mixture model, without the hydrostatic pressure assumption, that can output and analyze the data of gas concentration, water temperature, mixture velocity and turbulence parameters; the first to demonstrate, in the form of a temperature nephogram, the details of temperature change at the laboratory scale; the first to analyze the extent to which bubble plumes can affect destratification, which is very important for the designer of a bubble plume system to select an optimum aeration condition.

## 2 Mathematical model and numerical implementation

### 2.1 Mixture model

Models for bubbly flows have been generally divided into EL, (Sheng and Irons, 1995), and EE models (Bernard et al., 2000; Buscaglia et al., 2002; Mudde and Simonin, 1999; Sokolichin and Eigenberger, 1995; Smith, 1998; Yang et al., 2019). In this study, the two-phase fluid model using a double-averaging approach derived by Buscaglia et al. (2002) is adopted. The first average is performed at spatial scales of the order of bubble-to-bubble spacing (and smaller), leading to the conservation equations of mass and momentum for an air-water mixture, while the second average is the Reynolds-averaging over the air-water mixture equations at larger turbulent scales. Therefore, the double-averaged Navier-Stokes equations of mass and momentum conservation for mixture fluid are as follows,

$$\nabla \cdot \mathbf{u_m} = 0, \tag{1}$$

$$\frac{\partial \mathbf{u_m}}{\partial t} + \mathbf{u_m} \cdot \nabla \mathbf{u_m} + \frac{1}{\rho_m} \nabla p_m = \nabla \cdot [2(v + v_t)\mathbf{S}] - g\mathbf{k}, \tag{2}$$

in which $\mathbf{u_m}$, $p_m$ and $\rho_m$ represent the mixture quantities of fluid velocity, pressure and density, respectively, $\mathbf{k}$ the vertical unit vector, $g$ the gravitational acceleration, $\mathbf{S}$ the rate of strain tensor of mean flow defined by

$$S = \frac{1}{2}(\nabla \boldsymbol{u_m} + \nabla^T \boldsymbol{u_m}), \tag{3}$$

$v$ the kinematic viscosity, and $v_t$ and eddy viscosity, which will be calculated by solving $k$-$\varepsilon$ equations. In eqn (2),

$$\rho_m = (1 - \alpha_b)\rho_w + \alpha_b\rho_a, \tag{4}$$

where $\alpha_b$ is the volume fraction of bubbles following the definition in Drew and Passion (1998), $\rho_a$ the density of air and $\rho_w$ the density of water, which is related to water temperature $T$

$$\rho_w = 999.842\,594 + 6.793\,952 \times 10^{-2}T - 9.095\,290 \times 10^{-3}T^2 + 1.001\,685 \times$$
$$10^{-4}T^3 - 1.120\,083 \times 10^{-6}T^4 + 6.536\,332 \times 10^{-9}T^5. \tag{5}$$

In NEWTANK, a two-step projection method is employed to solve the double-averaged Navier-Stokes equations, with the time derivative term discretized by forward time difference. The combination of central difference method and upwind method is used to discretize the convection terms, while only the central difference is adopted for the pressure gradient and the stress gradient terms.

## 2.2　The equation of air volume fraction

In this study, the air volume fraction equation derived by Buscaglia et al. (2002) is employed. The dissolution of air in water is neglected and the air bubble phase is considered to be a single, inert component. The equation of air/bubble volume fraction is given by

$$\frac{\partial \alpha_b}{\partial t} + \nabla \cdot \left( \alpha_b \mathbf{u_g} \right) = \nabla \cdot \left( D_g \nabla \alpha_b \right), \tag{6}$$

where $\mathbf{u_g}$ is the bubble advection velocity that can be calculated by

$$\mathbf{u_g} = \mathbf{u_m} + w_s \mathbf{k}, \tag{7}$$

in which $w_s$ is the bubble-slip velocity assumed to depend on the bubble radius (Clift et al., 1978),

$$w_s = \begin{cases} 4474 \text{m/s} \times r_b^{1.357} & if\ 0 \leq r_b \leq 7 \times 10^{-4} \text{ m} \\ 0.23 \text{m/s} & if\ 7 \times 10^{-4} \leq r_b \leq 5.1 \times 10^{-3} \text{ m}, \\ 4.202 \text{m/s} \times r_b^{0.547} & if\ r_b > 5.1 \times 10^{-3} \text{ m} \end{cases} \tag{8}$$

$D_g$ the dispersion coefficient associated with the turbulence and bubble-bubble interaction, and according to Carrica et al. (1998),

$$D_g = \frac{v_t}{S_g}, \tag{9}$$

where $S_g$ is the Schmidt number for air, and the value of $S_g$ is taken as 1.0.

In Eq. (6), the time derivative term is discretized by the forward time difference method. Given that $\mathbf{u_g}$ consists of $\mathbf{u_m}$ and $w_s \mathbf{k}$, the convection terms are divided into mixture convection terms and bubble-slip terms. Eq. (6) is solved by two steps. In the first step, an intermediate air concentration is introduced and only bubble-slip terms and the diffusion terms are included. In the second step, the convection terms of air concentration are calculated by using an approach similar to the VOF method adopted in Liu (2007). Thus, the movement of bubbles and the track of the free surface can be unified and achieved simultaneously by this method. For two-phase fluid, when $0 < \alpha_b < 1.0$, the grids may be located at the water-air interface or below the interface when bubble plumes exist. However, in this paper, the free surface reconstruction is only carried out for the girds located at the water-air interface. If the grid with the value of $\alpha_b$ between 0 and 1.0 is located below the free surface, only upwind scheme will be adopted for the convection of bubble concentration.

## 2.3 Turbulence model

In this study, the $k$-$\varepsilon$ model is employed to model the turbulent dissipation, with effects of aeration and water temperature on turbulence, i.e.,

$$\frac{\partial k}{\partial t} + \nabla \cdot (k\mathbf{u_m}) = \nabla \cdot \left[\left(\nu + \frac{\nu_t}{\sigma_k}\right)\nabla k\right] + \nu_t |\mathbf{S}|^2 - \varepsilon - \frac{\nu_t}{Pr_t}\frac{1}{\rho_0}g\frac{\partial \rho_m}{\partial x}, \tag{10}$$

$$\frac{\partial \varepsilon}{\partial t} + \nabla \cdot (\varepsilon \mathbf{u_m}) = \nabla \cdot \left[\left(\nu + \frac{\nu_t}{\sigma_\varepsilon}\right)\nabla \varepsilon\right] + C_{1\varepsilon}\nu_t |\mathbf{S}|^2 \frac{\varepsilon}{k} - C_{2\varepsilon}\frac{\varepsilon^2}{k} - C_{3\varepsilon}(\frac{\nu_t}{Pr_t}\frac{1}{\rho_0}g\frac{\partial \rho_m}{\partial x}), \tag{11}$$

$$\nu_t = C_d \frac{k^2}{\varepsilon}, \tag{12}$$

where $Pr_t$ is turbulence Prandtl number, with the value of 0.85; $C_d$, $\sigma_k$, $\sigma_\varepsilon$, $C_{1\varepsilon}$, $C_{2\varepsilon}$ and $C_{3\varepsilon}$ are empirical coefficients with recommended values as

$$C_d = 0.09, \ \sigma_k = 1.0, \ \sigma_\varepsilon = 1.0, \ C_{1\varepsilon} = 1.44, \ C_{2\varepsilon} = 1.92, \ C_{3\varepsilon} = 1.44$$

The last term in Eq. (10) represents the generation of turbulence due to buoyancy and it reflects the effects of aeration and water temperature since $\rho_m$ varies with changes of gas concentration and water density related to water temperature. $C_{3\varepsilon}$ is the coefficient of the buoyancy production term which represents the degree to which turbulence dissipation rate $\varepsilon$ is affected by the buoyancy. In this study, $C_{3\varepsilon} = 1.44$ is adopted, as suggested by Rodi (1980).

## 2.4 Energy model

The energy equation for the two-phase flow is as follows,

$$\frac{\partial}{\partial t}\left(\sum(\alpha_l\rho_lC_{pl}T_m)\right) + \nabla \cdot \left(\sum \alpha_l\mathbf{u}_l\rho_l(C_{pl}T_m + p)\right) = \nabla \cdot (k_{eff}\nabla T) \tag{13}$$

where $T_m$ represents the temperature of mixture fluid, $\alpha_l$, $\rho_l$, $C_{pl}$ and $u_l$ respectively represent the volume fraction, density, specific heat and velocity of phase $l$. $l$=1 means water phase and $l$=2 means air bubble phase:

$$\alpha_1 = 1.0 - \alpha_b, \rho_1 = \rho_w, C_{p1} = C_{pw}, \mathbf{u_1} = \mathbf{u_m}$$

$$\alpha_2 = \alpha_b, \rho_2 = \rho_a, C_{p2} = C_{pa}, \mathbf{u_2} = \mathbf{u_g}$$

$k_{eff}$ is the effective conductivity:

$$k_{eff} = \lambda + \rho_m \frac{C_p\nu_t}{Pr_t} \tag{14}$$

in which $C_p$ and $\lambda$ respectively represents specific heat and conductivity of mixture fluid, so $C_p$ and $\lambda$ can be updated as,

$$C_p = (1 - \alpha_b)C_{pw} + \alpha_b C_{pa} \tag{15}$$

$$\lambda = (1 - \alpha_b)\lambda_w + \alpha_b \lambda_a \tag{16}$$

In Eq. (13), the time derivative terms are discretized in the forward difference method. The first-order upwind difference scheme is employed to solve the convective terms, and the diffusion terms are discretized by the central difference method.

## 3 Model verifications

In this section, the proposed numerical model is validated against the experimental data by Zarrati (1994) and Zic (1990).

### 3.1 Bubble plume in an open channel

In the experiment by Zarrati (1994), air was injected from the bed of an open channel and the air concentration profiles were measured at different sections downstream. the detailed experimental setup is shown in Fig. 1. The channel is 15.0 cm wide with a slope of 14.5°. The sluice gate opening was 3.5 cm, and the water entered a flow channel in 3.4 m/s. An air diffuser was installed at the bed 1.5 m downstream of the gate. Needle probes were installed at several sections away from the aeration position to measure the air concentration. Measurements showed that for the air volume fraction greater than 5%, most of bubbles were greater than 2mm.

In the numerical simulation, the *x*-direction of coordinate system is set along the open channel, so the gravitational acceleration in *x*- and *z*-direction is $g_x = g\sin\theta$ and $g_z = g\cos\theta$. The employed grid system has 800 uniform grids with the grid size $\Delta x = 5$ mm in *x*-direction and 45 uniform grids with the grid size $\Delta z = 1$ mm in *z*-direction. In this case, the aeration rate was set to be 1.0 m/s and the bubble-slip velocity $w_s$ was taken as 0.23 m/s.

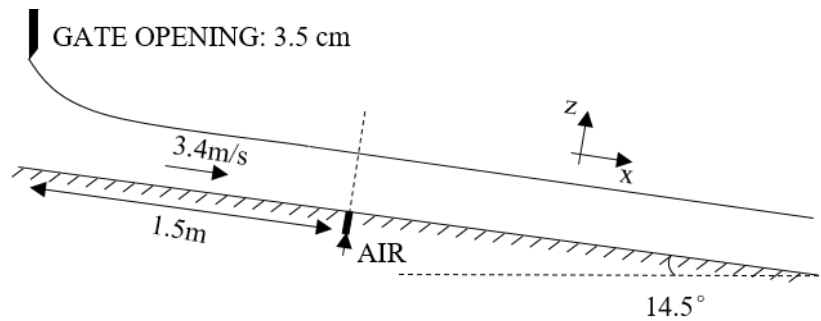

Fig. 1 The open channel and the setup in the experiment.

The spatial distributions of air concentration and eddy viscosity are shown in Fig. 2. It can be seen from Fig. 2(a) that after released from the open gate, the water depth decreases gradually but when crossing the position of bubble plume releasing, the surface can be raised slightly under the influence of aeration. Meanwhile, after the bubble plumes reach the flow surface, surface aeration can be generated downstream, so the clear interface between air and water cannot be sustained any more. In addition, it is shown in Fig. 2(b) that downstream the open gate, the kinematic eddy viscosity $v_t$ increases gradually from the bed of channel and develops towards the water surface. Downstream of aeration position, the value of $v_t$ increases abruptly, which indicates that the turbulence inside water is enhanced by the bubble plumes from the channel bed.

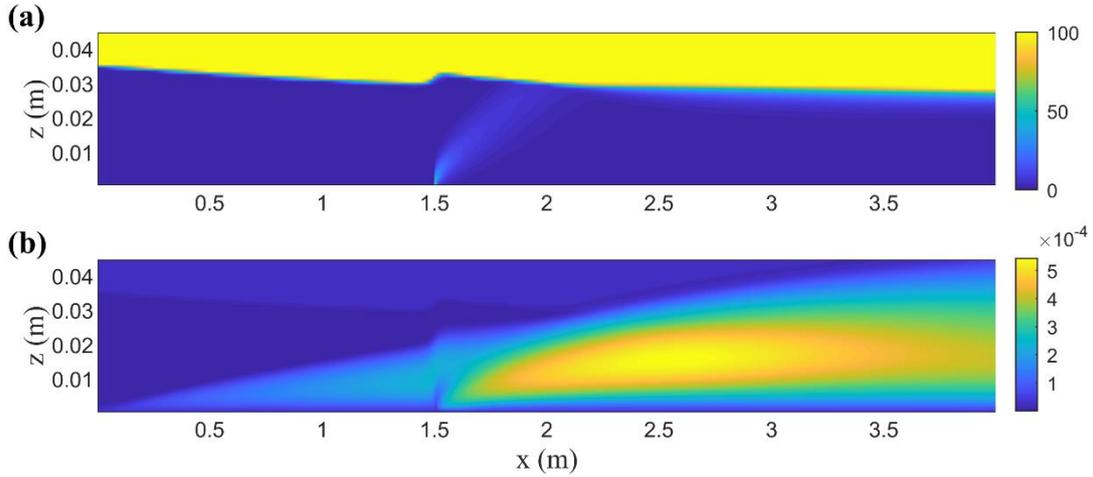

Fig. 2 Spatial distribution of (a) air concentration $\alpha_b$ (%) and (b) kinematic eddy viscosity $v_t$.

Furthermore, the comparison of air concentration profiles at the four downstream sections between experimental data and numerical results are shown in Fig. 3, from which it can be found that good agreement is observed, indicating that the accuracy of the mixture model and the tracking air concentration proposed numerical model are acceptable for simulating bubble plumes in open channel. When bubble plumes are released from the diffuser, they will rise rapidly under the effects of buoyancy. Meanwhile, affected by the water flow, the bubbles are carried downstream by the water flow and eventually escape to the surface. From 10 to 30 cm downstream of the aeration position, the gas concentrations near the channel bed, at $z$ = 0.25, 0.5 and 0.75 cm, rapidly decrease from 3.1%, 7.2% and 11.3% to 0.4%, 1.1% and 2.2%. Meanwhile, the gas concentration near the water surface, at $z$ = 1.75, 2.0 and 2.25 cm, increase from 3.0%, 1.3% and 0.4% to 7.5%, 7.3% and 6.1%. The section of the maximum gas concentration was gradually approaching the water surface, from $z = 0.95$ to $1.85$ cm, until bubbles get out of the water completely. Throughout this process,, the maximum gas concentration continuously decreased due to the action of convection and diffusion, from 12.7% to 7.5%. This demonstrates that the present model can simulate the development of air-water mixtures with satisfactory accuracy.

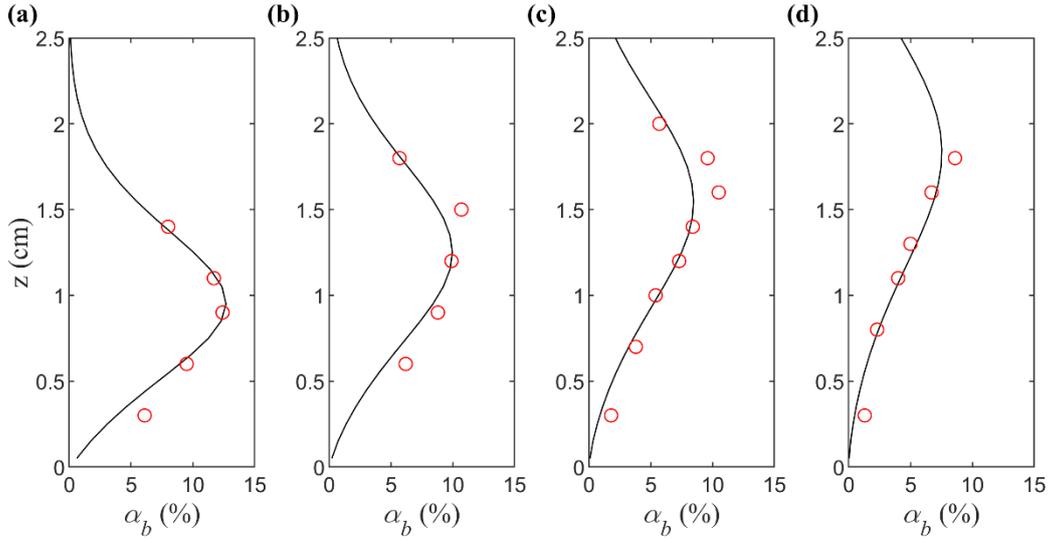

Fig. 3 Comparisons of air concentration profiles at the sections of (a) 10, (b) 15, (c) 20 and (d) 30 cm downstream of the aeration position between the experimental data (red circles) and numerical results (black solid line).

## 3.2 Destratification by bubble plumes

In this section, the experimental data and numerical results from Zic (1990) will be employed to further validate our numerical model. The experiments were performed in a 3D rectangular tank with dimensions of $1.1 \times 1.1 \times 1.4$ m (Fig. 4). Compressed air was injected through a diffuser 1.2 cm in radius and 3.5 cm in length. The diffuser was located in the center of the tank and the distance between the diffuser and the bottom can be adjusted. The initial stratification was achieved by filling the tank from the bottom with a prepared mixture of cold and heated water. Temperature profiles were measured by a vertical chain of 16 thermistors which are 0.3 m away from the plume axis (Fig. 4b) and were recorded every 20 s. The experiments typically lasted for 10 minutes until total mixing was completed. Water temperature measurements began with the initiation of air release ($t = 0$) and continued until the water body was well mixed.

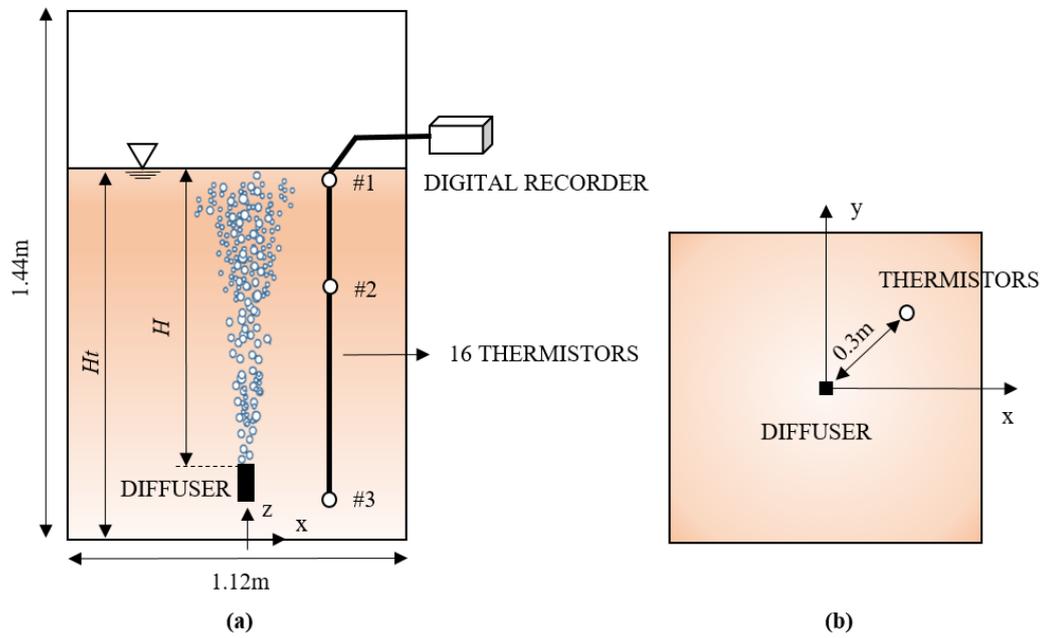

**(a)**                                    **(b)**

Fig. 4 The (a) front view and (b) top view of the rectangular tank utilized in the experiment; thermistors #1 and #2 represents those located 0.01 and 0.65 m below the free surface, and the thermistor #3 is at the same depth as the aeration diffuser (0.89 m for Case 1 and 0.97 m for Case 2).

Two cases in Zic (1990) are chosen in this study. Table 1 shows the parameters in the experiments, including aeration rate ($Q_{air}$), diffuser submergence depth ($H$), total water depth ($H_t$), and the initial (maximum) water temperature difference between surface and bottom ($\Delta T_{t=0}$).

Table 1 Conditions of the Experiments in Zic (1990).

| Case | $Q_{air}$ | $H$ | $H_t$ | $\Delta T_{t=0}$ | $Q_{air}/A$ |
|------|-----------|-----|-------|------------------|-------------|
|      | (Ls$^{-1}$) | (m) | (m) | (°C) | (Ls$^{-1}$m$^{-2}$) |
| 1 | 0.10 | 0.89 | 1.10 | 18.0 | 0.080 |
| 2 | 0.067 | 0.97 | 1.22 | 15.1 | 0.053 |

The initial water temperature profiles in the tank are shown in Fig. 5 according to Zic (1990).

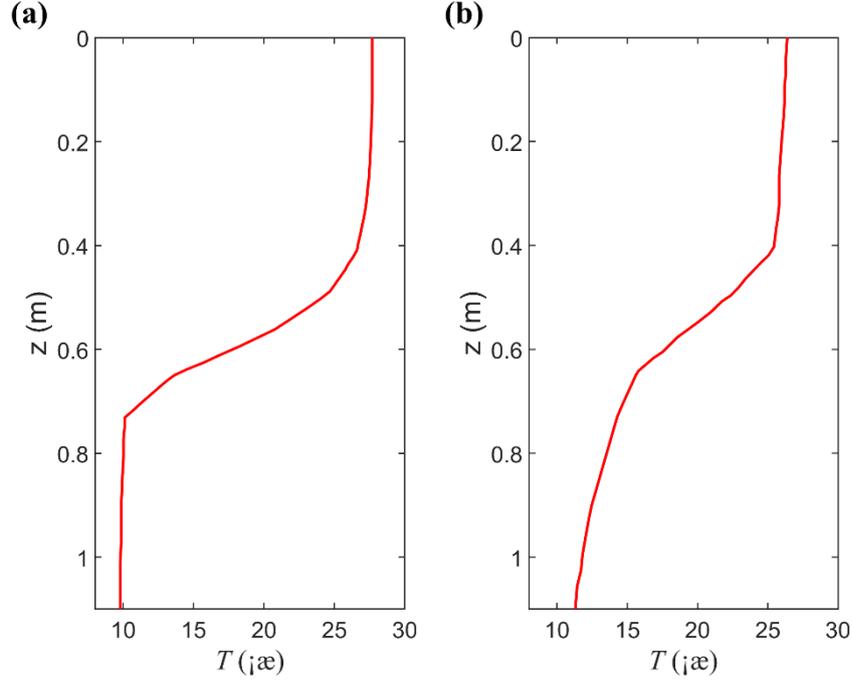

Fig. 5 The initial water temperature profiles for (a) Case 1 and (b) Case 2.

In numerical simulation, the employed mesh system has 57 non-uniform horizontal meshes with the minimum $\Delta x = \Delta y = 0.01$ m arranged near the aeration position in both $x$- and $y$-direction. In vertical direction, 26 non-uniform grids with mesh size $\Delta z = 0.06\text{-}0.07$ m are utilized. The size of the bubbles released during the experiment is in the range of 2-5 mm, so according to Eq. (8), the value of $w_s$ was taken as 0.23 m/s.

For Case 1, the distributions of temperature and the velocity field at the mid-plane of the reservoir at $t$ = 0, 8, 24, 36, 48, 80, 152, 200 and 328 s are shown in Fig. 6. It can be seen from Fig. 6(a) that obvious temperature stratification has been formed in the reservoir before aeration. Fig. 6(b) shows that with air released from the diffuser, the bubbles rise upwards to the free surface under the force of buoyancy and can carry ambient water with them, forming bubble plumes at $t$ = 8 s. When the bubbles reach the water surface, the carried cold water will spread outwards and mix with the epilimnion water, and thus the phenomenon can occur that the water at the surface is colder than that below the surface. It can be seen from Fig. 6(c) and 6(d) that when surface water spreads to the tank wall, it drops along the wall, and then under the action of water flow, the water with temperature 22-24℃ mixes with water at intermediate depth. Meanwhile, the thermocline keeps moving down and getting denser, and the water at 20-22℃ similarly drops along the wall. Fig. 6(e)-6(h) show that when the temperature of water at the intermediate depth reaches 22-24℃, the water of 20-22℃ is also mixed with it at the same time and the mixing region is becoming increasingly large. When the temperature of water at intermediate depth decreases to 20-22℃, the thermocline is still moving down and getting denser. This process repeats several cycles until the water in the tank is mixed completely at $t$ = 328 s, as is shown in Fig. 6(i).

The distribution of air concentration when the water with different temperature is completely mixed at $t$ = 800 s is shown in Fig. 7(a). The air is released 0.25 m away from the bottom and the maximum gas concentration appears near the location of diffuser. As bubbles rise and diffuse around, the air concentration decreases consequently. At $t$ = 800 s, the intensity of turbulence kinetic energy $k$ and mixture-fluid velocity $\mathbf{u}_m$ are also shown in Fig. 7(b). Since the aeration rate remains unchanged at the position of diffuser, the largest turbulence kinetic energy intensity appears above the air diffuser. During the rising of bubbles, the velocity and turbulence kinetic energy in the center of the plumes keep decreasing gradually, but the affected area by bubble plumes becomes larger. When bubble plumes reach surface, the turbulence kinetic energy becomes larger, probably because the bubbles are getting out of the water-air interface. When the flow propagates to the wall of tank, it can be seen from Fig. 7(b) that it dropped to form a circulation. During the whole circulation, the turbulence kinetic energy also decreased continuously.

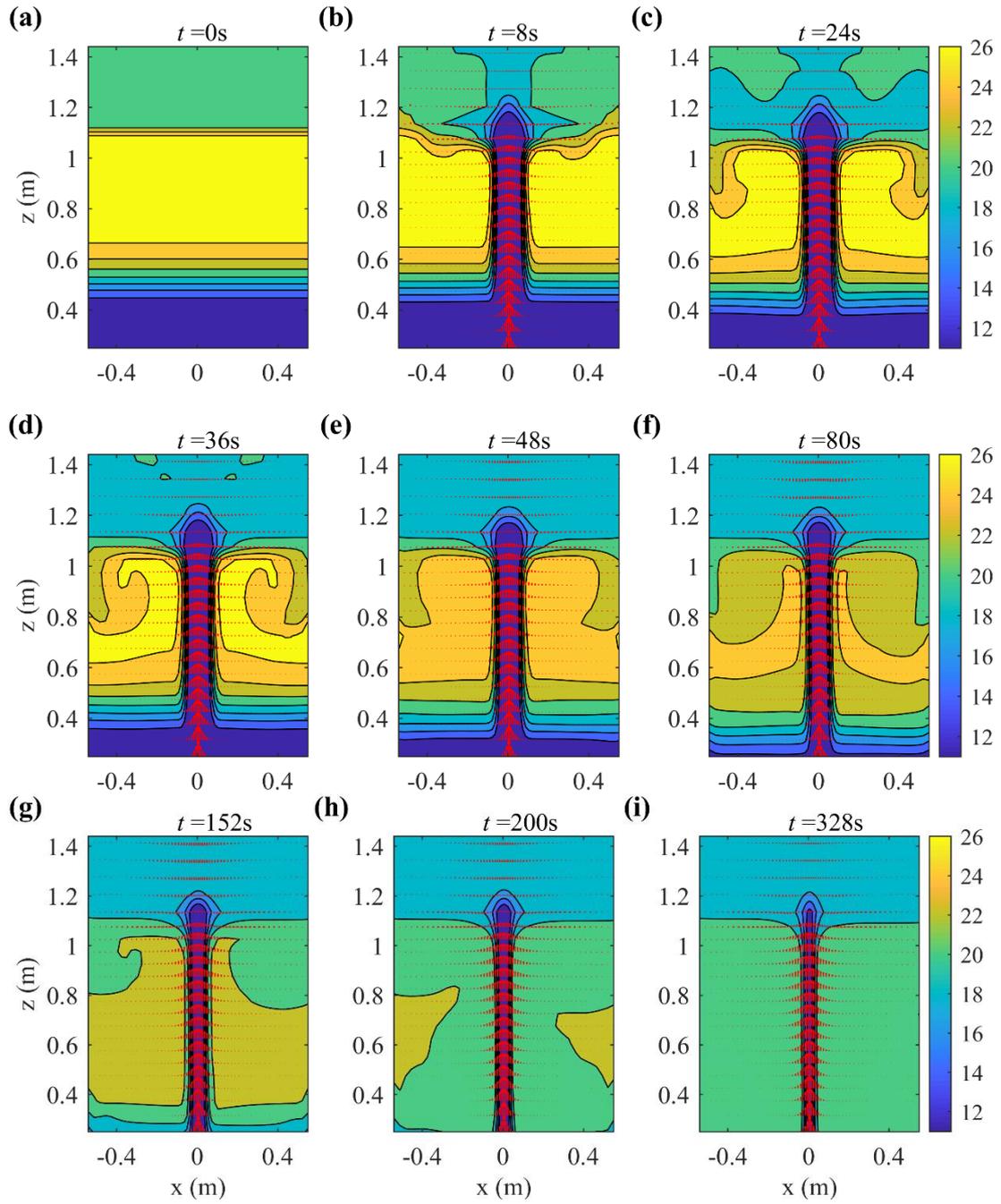

Fig. 6 Numerical results of water temperature structure with velocity field of fluid flow at $t =$ (a) 0, (b) 8, (c)24, (d) 36, (e) 48, (f) 80, (g) 152, (h) 200 and (i) 328 s in Case 1.

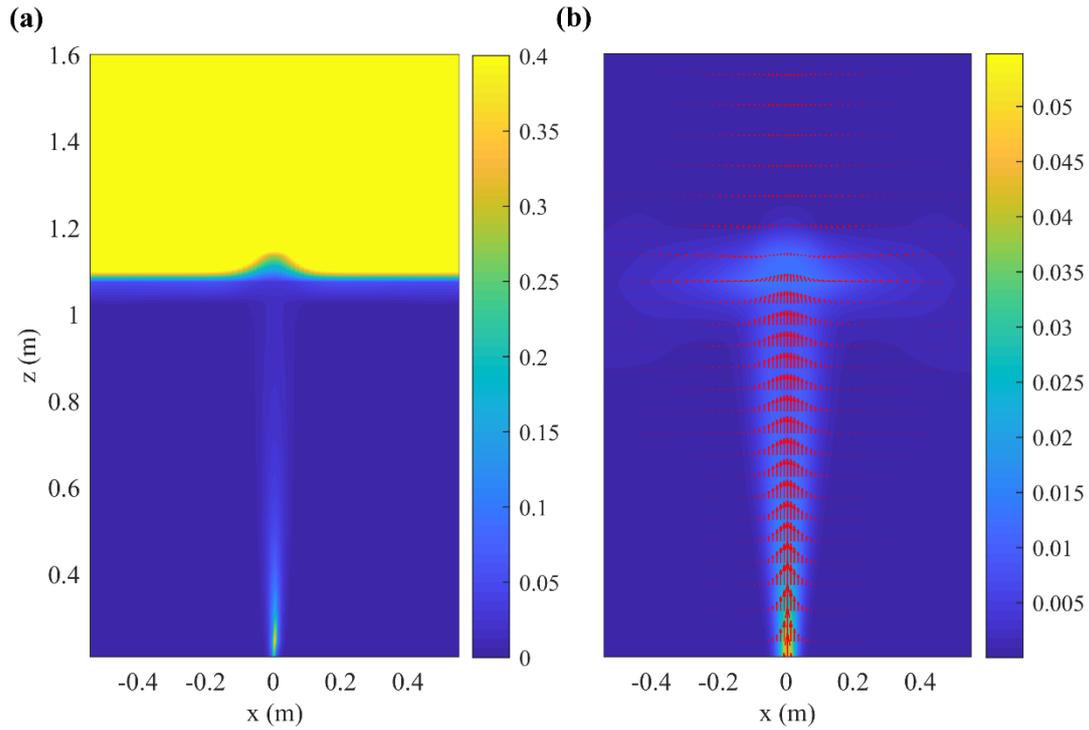

Fig. 7 (a) The distribution of air concentration $\alpha_b$ and (b) the distribution turbulence kinetic energy $k$ with velocity field of fluid flow at $t = 800$ s in Case 1.

Fig. 8 shows the comparison of the time series of water temperature at the position of #1, #2 and #3 (Fig. 4a) between the numerical results and experimental data in Zic (1990). It can be seen from Fig. 8 that at the beginning of aeration, the temperature near water surface (#1) decreases rapidly while the temperature at #2 starts to rise, because the bubble plumes carry cold water from the bottom to the surface and the cold water mixes with the surface water. Then the temperature at #1 gets equal to that at #2 at about $t = 80$ s, after which the water temperature at #2 becomes higher than that of #1 and then begins to drop, and finally the temperatures at #1 and #2 approach to the same value. As for the water temperature at #3, it increases slowly and takes a longer time to be equal to the temperature near free surface. This is probably due to the fact that the position of #3 is at the same depth to aeration diffuser and it would take much more time to feel the effects of bubble plumes. As is shown in Fig. 8, the water-temperatures at #1 and #2 match the numerical results better, and the intersection time is almost the same as that in the experiment. It can be also seen in Fig. 9 that the predicted temperature profiles at different time are very closer to the experimental data, and the evolution of the temperature profiles is highly consistent with the previous results in Singleton et al. (2010) and Chen et al. (2017). However, it can be seen from both Fig. 8 and Fig. 9 that the temperature at #3 in numerical simulation changes faster and the thermocline thickness is larger than the experimental data. It may result from the fact that the coalescence, breakup and collisions of bubbles have not been considered in this model, which may reduce the turbulent intensity, so the mixture of temperature at #3 is faster in the numerical simulation than that in the experiment.

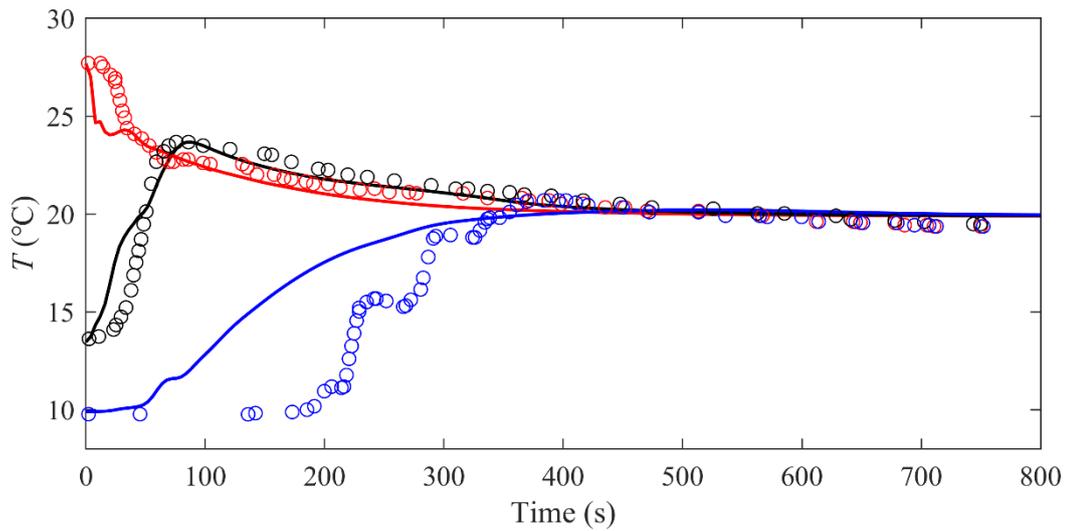

Fig. 8 Comparisons of time series of water temperature at the position of #1 (red), #2 (black) and #3 (blue) among the present numerical results (solid line) and experiment data (circle) in Case 1.

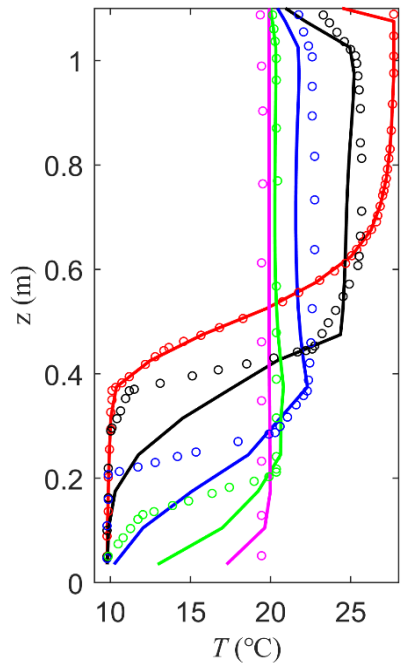

Fig. 9 Comparisons of temperature profiles at $t$ = 0 (red), 61 (black), 182 (blue), 365 (green) and 754s (magenta) between the present numerical results (solid line) and experiment data (circle) in Case 1.

The horizontal distributions of vertical velocity and air concentration at different depths are shown in Figs. 10 and 11, respectively. The further away from the aeration position, the smaller the gas concentration at centerline significantly gets, and the larger the bubble area becomes due to the diffusion action. Correspondingly, the mixture velocity at centerline has a slight decrease and the range of mixture velocity gets wider. From $t = 200 \text{ s}$ when water temperature is still stratified to $t = 800 \text{ s}$ when water temperature is mixed completely, the distributions of mixture velocity and air concentration at different sections have no obvious change. Even though the water temperature has changed dramatically, the bubble plume will not be affected significantly as long as it reaches stability. It can be seen from the Fig. 10 that the numerical model predicts well the self-similarity of mixture velocity along different downstream sections from $t = 400$ to 800 s. It can be seen from Fig. 10 that the mixture velocity at $t = 200$ s does not present self-similarity, probably because the temperature structure is still stratified at this moment. Fig. 11 shows that from $t = 200$ to 800 s, the air concentration along different sections all presents good self-similarity, because the change of water temperature has no significant effect on the rise, advection and diffusion of bubbles.

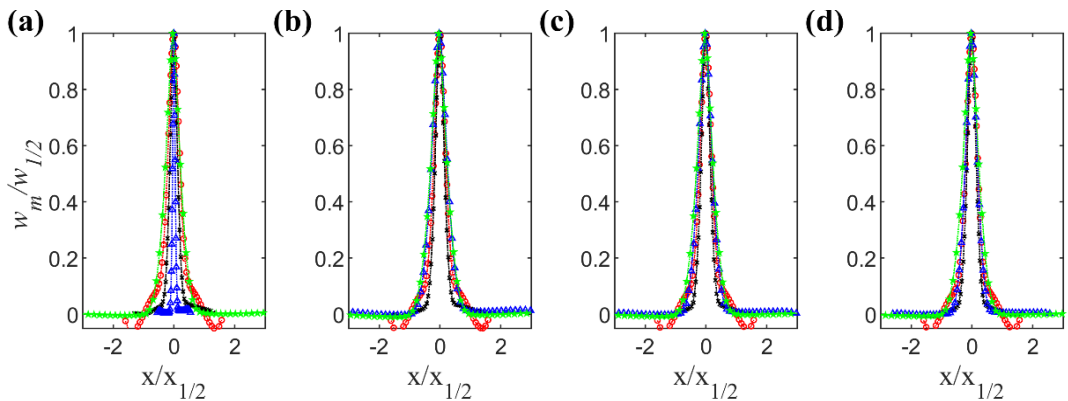

Fig. 10 Horizontal distribution of normalized mixture velocity $w_m/w_s$ at different depths away from the surface, i.e., 0.23 (dot-circle line), 0.47 (dot-cross line), 0.65 (dot-triangle line) and 0.83 m (dot-pentagram line), at $t =$ (a) 200, (b) 400, (c) 600 and (d) 800 s, in Case 1. $w_{1/2}$ is the mixture fluid velocity at the centerline and $x_{1/2}$ is the half-width of the bubble plume.

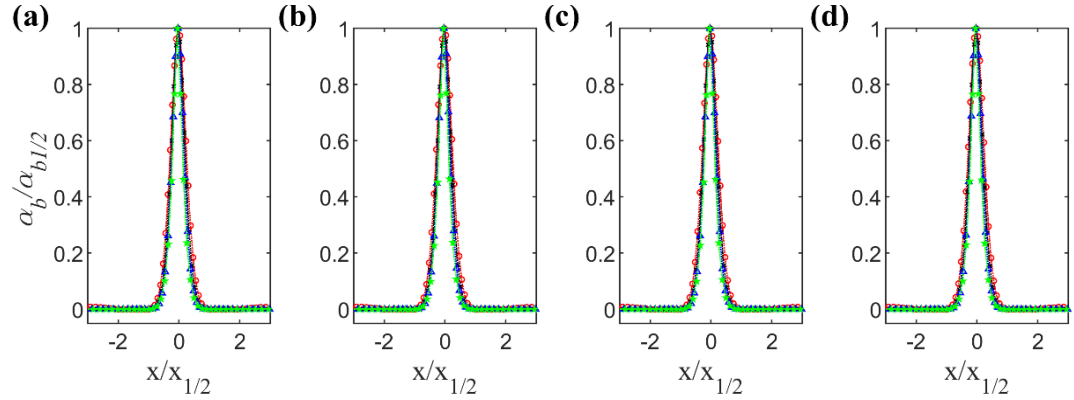

Fig. 11 Normalized gas concentration along different cross sections at depth of 0.23 m (dot-circle line), 0.47 m (dot-cross line), 0.65 m (dot-triangle line) and 0.83 m (dot-pentagram line), when $t =$ (a) 200, (b) 400, (c) 600 and (d) 800 s, in Case 1. $\alpha_{b1/2}$ is the gas concentration at the centerline and $x_{1/2}$ is the half-width of the bubble plume.

As for Case 2, Fig. 12 shows the time series of water temperature at the position of #1, #2 and #3. Both the experimental data and numerical results from Zic (1990) are compared with the numerical results in this study. It can be seen from Fig. 12 that the numerical results of Zic (1990) show faster changes in the temperature field than the experimental data and the final stable temperature is also lower than experimental results. The numerical model adopted by Zic (1990) was actually a two-dimensional model, and according to his view, this mismatch may be related to the fact that the mixing is essentially a 3D problem which may produce additional damping effects. In this study, a 3D model was adopted, and the calculated temperature change and final stable temperature are much closer to the experimental results as shown in Fig. 12.

From Fig. 6 to Fig. 12, it can be seen that the numerical results of the present model are in good agreement with the dynamics of destratification induced by bubble plumes, indicating that the numerical model established in this paper can be utilized to predict the development and spatial distributions of such physical parameters as gas concentration, flow velocity and turbulence kinetic energy.

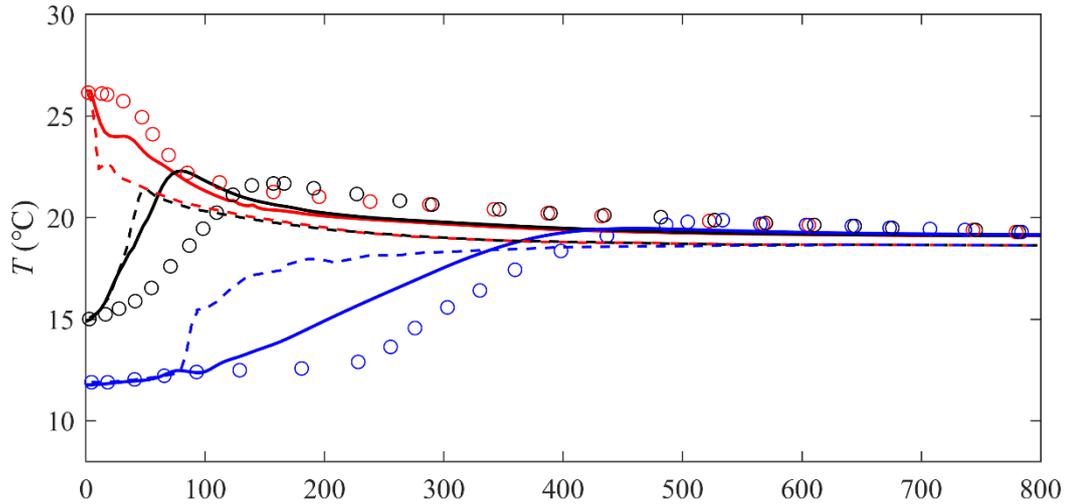

Fig. 12 Comparisons of time series of water temperature at the position of #1 (red), #2 (black) and #3 (blue) among the present numerical results (solid line), experiment data (circle) and numerical results from Zic (1990) (dash line) in Case 2.

# 4   Impact factors of bubble plumes on temperature destratification

In this section, the validated numerical model will be used to further study the impact factors of destratification induced by bubble plumes in idealized reservoirs, including the aeration rates, the aeration location, and the water area, based on the experimental setup of Case 2 in Section 3.2.

## 4.1   No application of bubble plumes

Before the investigation of bubble plumes, how temperature changes under the condition of no aeration in the reservoir is first simulated for comparison. It can be seen from Fig. 13 that basically the temperature structure keeps stable in 800 s. The subtle change and fluctuation are probably due to the molecular diffusion effects. In addition, Fig. 14 shows that compared with temperature profile at $t$ = 0 s, the stratified structure of water temperature could still maintain the profile in 800s. As a result, it is reasonable to believe that when bubble plumes are applied, the effects of destratification are mainly caused by aeration.

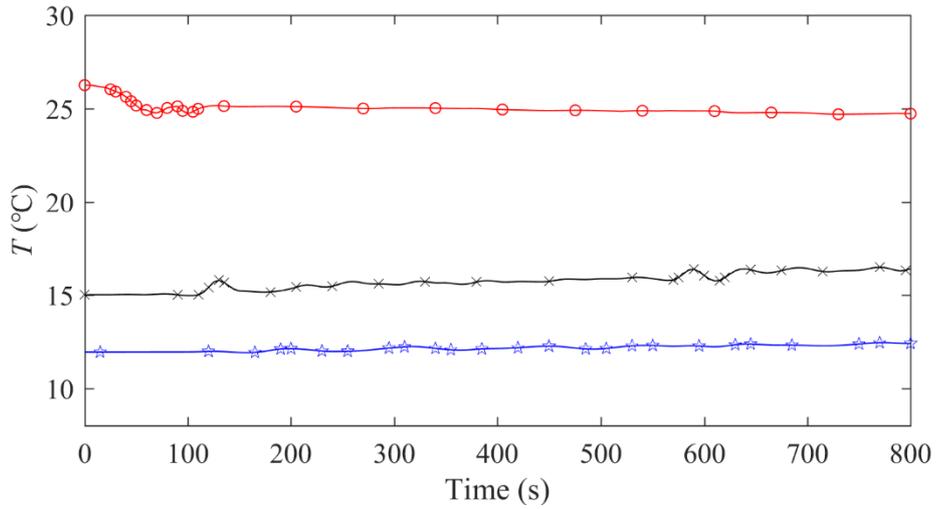

Fig. 13 Time series of water temperature of #1 (solid-circle line), #2 (solid-cross line) and #3 (solid-pentagram line), when no aeration is applied.

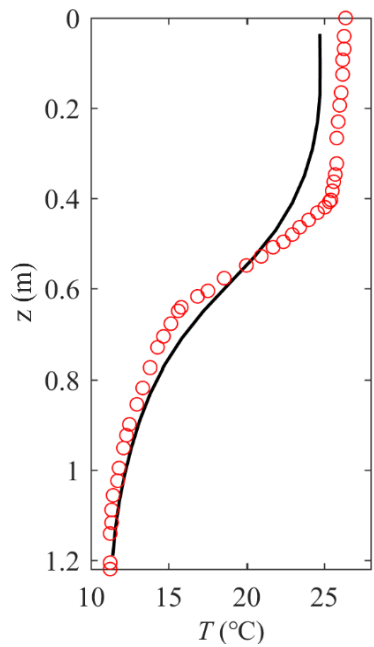

Fig. 14 Comparison of temperature profiles between numerical results when $t = 800\ s$ (solid line) and the initial temperature condition (circle) when no aeration is applied.

## 4.2 Impact of the aeration rate on destratification

For the same stratified structure, the destratification effects and efficiency of different aeration rates may change significantly. Asaeda and Imberger (1993) have studied the structure of bubble plumes in linearly stratified environment. Depending on the aeration rate with respect to stratification, he defined three types of intrusions for destratification, i.e., a single intrusion, a stack of intrusions and unsteady intrusions. It suggested that a large aeration rate, which usually generates only one intrusion, should be adopted. However, large aeration rate usually means more energy to be consumed. Therefore, the optimal aeration rate should be analyzed and studied. In this part, the effects of different aeration rates on destratification are investigated. In the simulation, the aeration location is fixed at 0.97 m below free surface, and a series of aeration rates ranging from 0.2-2.0 m/s, at an interval of 0.2 m/s, are used.

When the aeration rate was as small as 0.2 m/s, it is found that the water temperature fluctuates significantly and it takes more than 1000 s for the temperature to get stable, indicating that very small aeration rate is not efficient or suitable for destratification. Imteaz and Asaeda (2000) also found that the insufficient gas flow rate cannot realize the breakup of the thermocline. Larger than 0.2 m/s, the corresponding time needed for temperature mixing is quantitatively analyzed. In this study, the water temperature is considered to be well-mixed and stable when the absolute values of temperature differences among #1, #2 and #3 are less than 0.2 ℃. Fig. 15 shows that the time needed for the water to become well mixed under the condition of different aeration rates. The time needed for the temperature difference between #1 and #3, and between #1 and #2 less than 0.2 ℃ is also drawn. It can be seen from Fig. 15 that when the aeration rate is less than 1.4 m/s, with the increase of aeration rate, it takes less time for the water to be well-mixed.

However, on the other hand, when the aeration rate is larger than 1.4 m/s, even more time is needed to achieve the destratification. The reason why water needs more time to get well-mixed may be as follows. When aeration rate is larger than 1.4 m/s, it can be seen from Fig. 15 that the temperature at #3 actually needs more time than #2 to get stable. With a relatively low aeration rate, the temperature at #3 was smaller than that at #1 and would keep increasing to reach the stable temperature finally. However, when the aeration rate is too large, after the temperature at #3 is equal to that at #1, it will continue to become larger than that at #1, so it will take longer time for temperature to get well-mixed. Therefore, the optimal aeration rate should be around1.4 m/s in the setup chosen to avoid both long duration of releasing bubble plumes and high cost of energy consumed.

To determine the optimum aeration rate for achieving the fastest water mixing, a non-dimensional number $N_t$ is introduced as follows:

$$N_t = \frac{Q_a g}{6\pi A N^3 H_t \theta}$$

where $Q_a$ is the aeration rate at bottom of water tank, $A$ is the water surface area, $N$ is buoyancy frequency given by $N = \sqrt{-\frac{g}{\rho}\left(\frac{\partial \rho}{\partial z}\right)}$, $H_t$ is the water depth and $\theta = \frac{\Delta T_{well-mixed}}{\Delta T_{t=0}}$. Here, $\Delta T_{well-mixed}$ represents the temperature difference in well-mixed water. In this study, $\Delta T_{well-mixed}$ is set to 0.2℃.

When the value of $N_t$ approaches 1.0, the bubble plume achieves faster mixing of the stratified water. As shown in Fig. 16, when the $N_t$ equals to 1.0, the corresponding aeration rate is approximately 1.4 m/s.

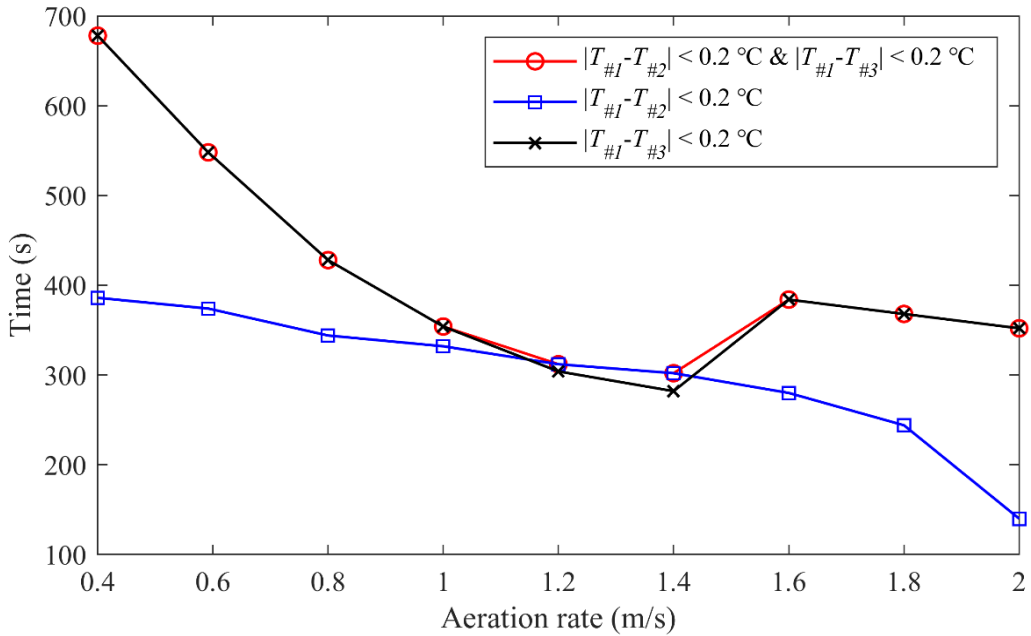

Fig. 15 Times needed for temperature stabilization under different aeration rates; ─□─: the time needed when the absolute temperature difference between #2 and #1 is less than 0.2 ℃, ─✕─ : the time needed when the absolute temperature difference between #3 and #1 is less than 0.2 ℃, and ─○─ : the time needed when both of the absolute temperature differences between #2, #3 and #1 are less than 0.2 ℃.

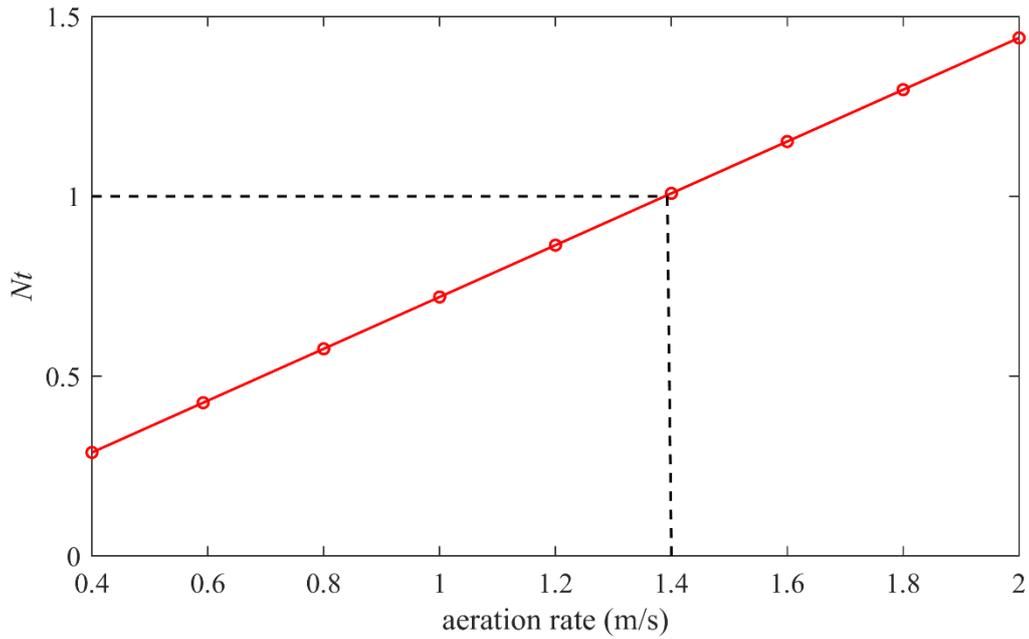

Fig. 16 $N_t$ under different aeration rates.

### 4.3 Impact of aeration location on destratification

The bubble plumes released from air diffuser can carry ambient cold water and rise to water surface, so the selection of aeration location (or diffuser location), from which bubbles start to rise, is very important. Given that the thermocline is about 0.6 m away from the tank bottom and the aeration location is generally lower than thermocline (Hughes et al., 1975), the aeration locations were chosen as 0, 0.12, 0.18, 0.25, 0.36, 0.48 and 0.60 m away from the tank bottom, with the aeration rate fixed to be 0.592 m/s.

Fig. 17 shows that generally as the aeration location is raised up, the time needed for temperature stabilization becomes longer. However, when the aeration location is higher than 0.24 m, the time needed for temperature stabilization does not change significantly. Therefore, the selection of aeration location should be set as close to the bottom of reservoir or lake as possible if the practical application condition is allowed. This is because when the aeration location is near the bottom, the bubble-induced convection can influence larger water body in vertical direction.

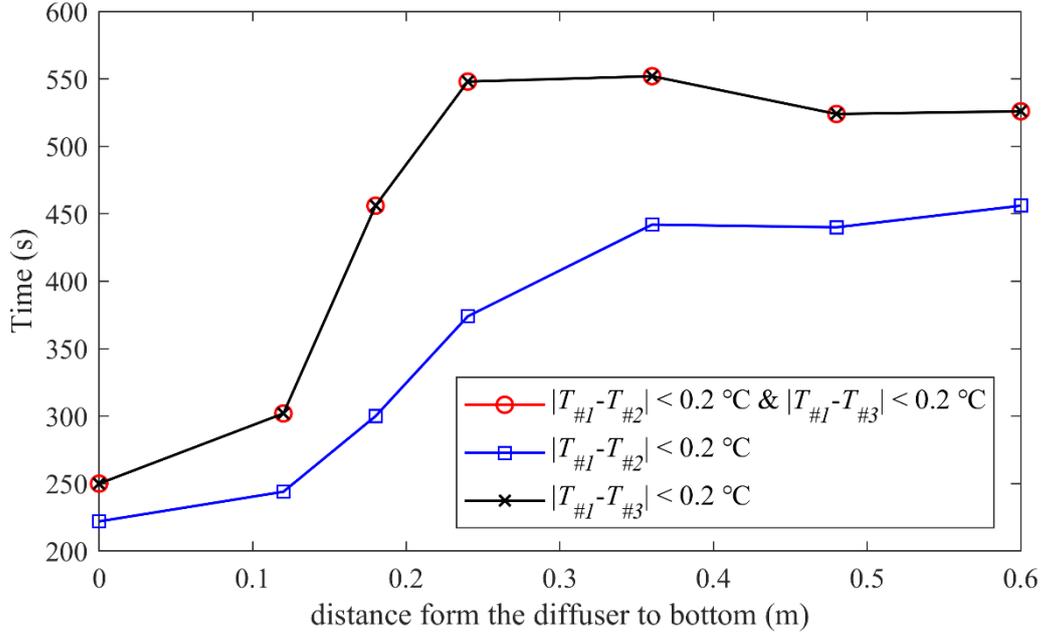

Fig. 17 The time needed for temperature stabilization at different aeration locations. The symbols are the same as those in Fig. 15.

### 4.4 Impact of the water area on destratification

#### 4.4.1 Destratification in a large water area

In previous case, the intrusion is formed along the tank wall and a vortex is developed between the intrusion and bubble plume. However, when practical reservoirs or lakes are considered, the water area may be so large that the intrusion is formed away from the water boundary and two symmetric vortexes are developed on both sides. Therefore, the mechanism of destratification in a larger water area is studied in this section. Due to the symmetric property, only half of the water area is simulated as is shown in Fig. 18. The air diffuser is located at the water tank bottom, with an aeration rate of 0.01 m/s and a water depth of 0.7 m. The initial temperature of the stratified water is set with a ratio of 0.7:1.22 according to Case 2.

In numerical simulation, the mesh system consists of 90 non-uniform horizontal meshes, with the minimum grid spacing $\Delta x = 0.01$ m located near the aeration position. In vertical direction, 50 uniform grids with a mesh size $\Delta z = 0.02$ m are utilized.

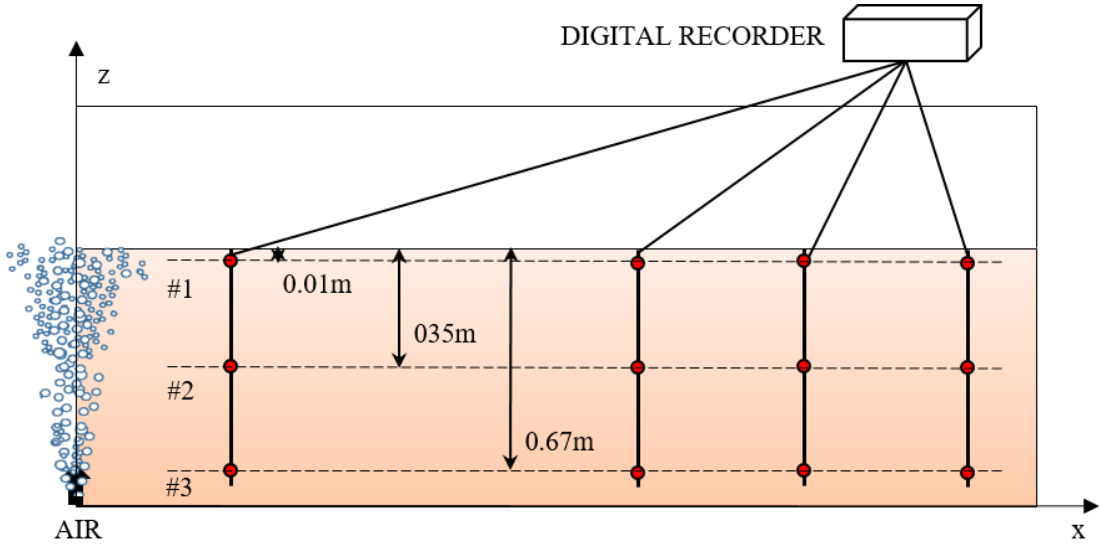

Fig. 18 The numerical setup of large water area with symmetric boundary condition.

The initial water temperature stratification of water area is shown in Fig. 19(a). After the aeration, as can be seen from Fig. 19(b), when the surface water spreads outwards for a short distance, it starts to drop and with an intrusion formed, a small vortex is developed near the water surface at $t = 8$ s. After a while at $t = 40$ s, it can be seen from Fig. 19(c) that as surface water keeps spreading, the first intrusion reaches a greater depth and meanwhile a second intrusion is formed. Fig. 19(d) shows that when the first intrusion reaches the level of neutral buoyancy, the second intrusion combines with the first intrusion, forms a large vortex and reaches to a larger water depth in Fig. 19(e). After that, it can be seen from Fig. 19(e)-(g) that two vortexes in opposite directions are developed on both sides of the intrusion and the area of inner vortex is getting increasingly small. The strongest inner vortex is form at $t = 160$ s in Fig. 19(g) and covers an aera of about 0.7m away from the aeration location. The length of the strongest inner vortex is very close to the plunging radius $R_s \sim \left(\frac{Q_0 g}{N^3}\right)^{\frac{1}{4}}$ by Lemckert and Imberger (1993). Due to the fact that the flow velocity of inner vortex is larger and that the inner vortex is surrounded by cold water, the water in inner vortex mixes faster and finally reaches a stable temperature, as shown in Fig. 19(h). After the inner water gets mixed completely, the area that has reached the stable temperature begins to develop outwards, but this process is very slow. It takes nearly 600 s to extend the stable mixing area from 1.0 m to 1.3 m as shown in Fig. 19(i), and about 1200 s from 1.3 m to 1.9 m in Fig. 19(j). Obviously, in actual engineering project, the mixing can be considered effective and efficient before 800 s. Therefore, if the water range with time needed for temperature stabilization in 800 s is considered to be the influence area of single bubble plumes, it can be seen that the single bubble plumes in this aeration condition can influence no more than 1.3 m. The influence area of single bubble plumes in different water areas is discussed in next section.

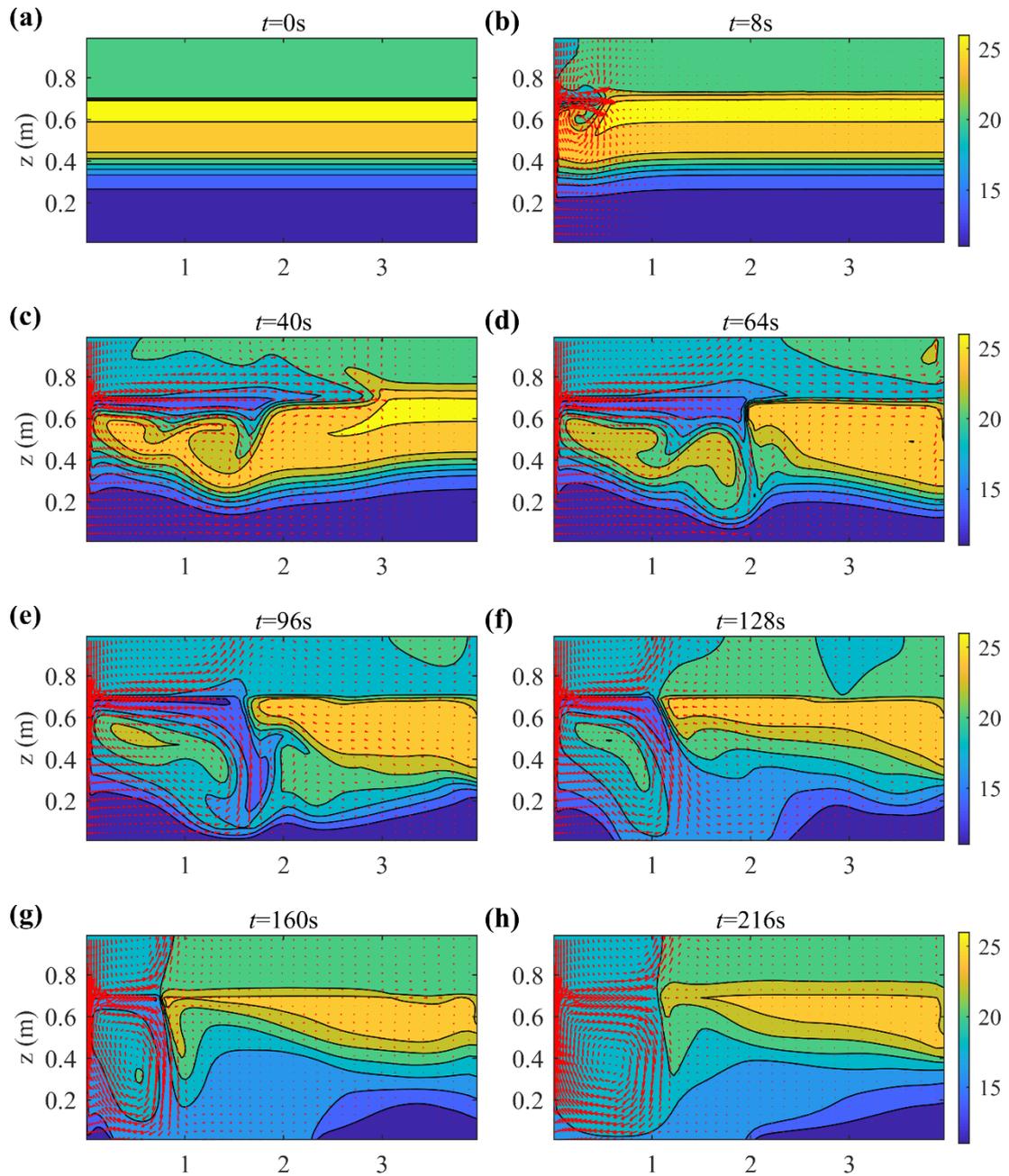

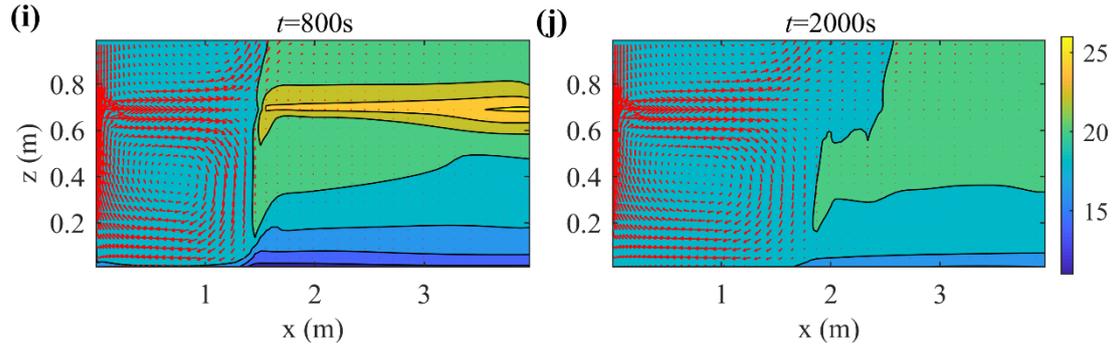

Fig. 19 The water temperature structure with velocity field of fluid flow at $t=$ (a) 0, (b) 8, (c) 40, (d) 64, (e) 96, (f) 128, (g) 160, (h) 216, (i) 800 and (j) 2000 s.

### 4.4.2 The influence area in different water areas

In this part, the setup is the same as the one in previous case except the water area. The influence of water area on destratification is studied by changing the half length of water area from 1.0 to 9.0 m.

It can be seen from Fig. 20(a) and 20(b) that when $x \leq 2.0$ m, the intrusion is formed at or very close to the right boundary, so the development of water temperature is similar to that shown in Fig. 6. Fig. 20(c)-(i) show that when $x \geq 2.5$ m, the intrusion is formed away from the right boundary, approximately at $x = 1.5$ m and the development trend of water temperature is nearly the same as that shown in Fig. 19. It is already known from Fig. 19 that the whole water area can hardly get mixed completely and the water of inner vortex mixes much faster than the outer one. As a result, seven sections at $x$=0, 0.2, 0.4, 0.6, 0.8, 1.0, 1.2 and 1.4 m are taken and the time needed for temperature stabilization at different sections are compared to confirm the influence area of single bubble plumes.

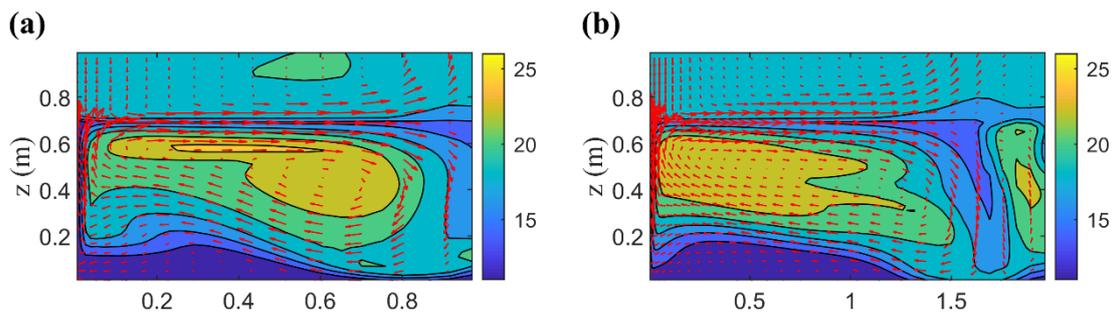

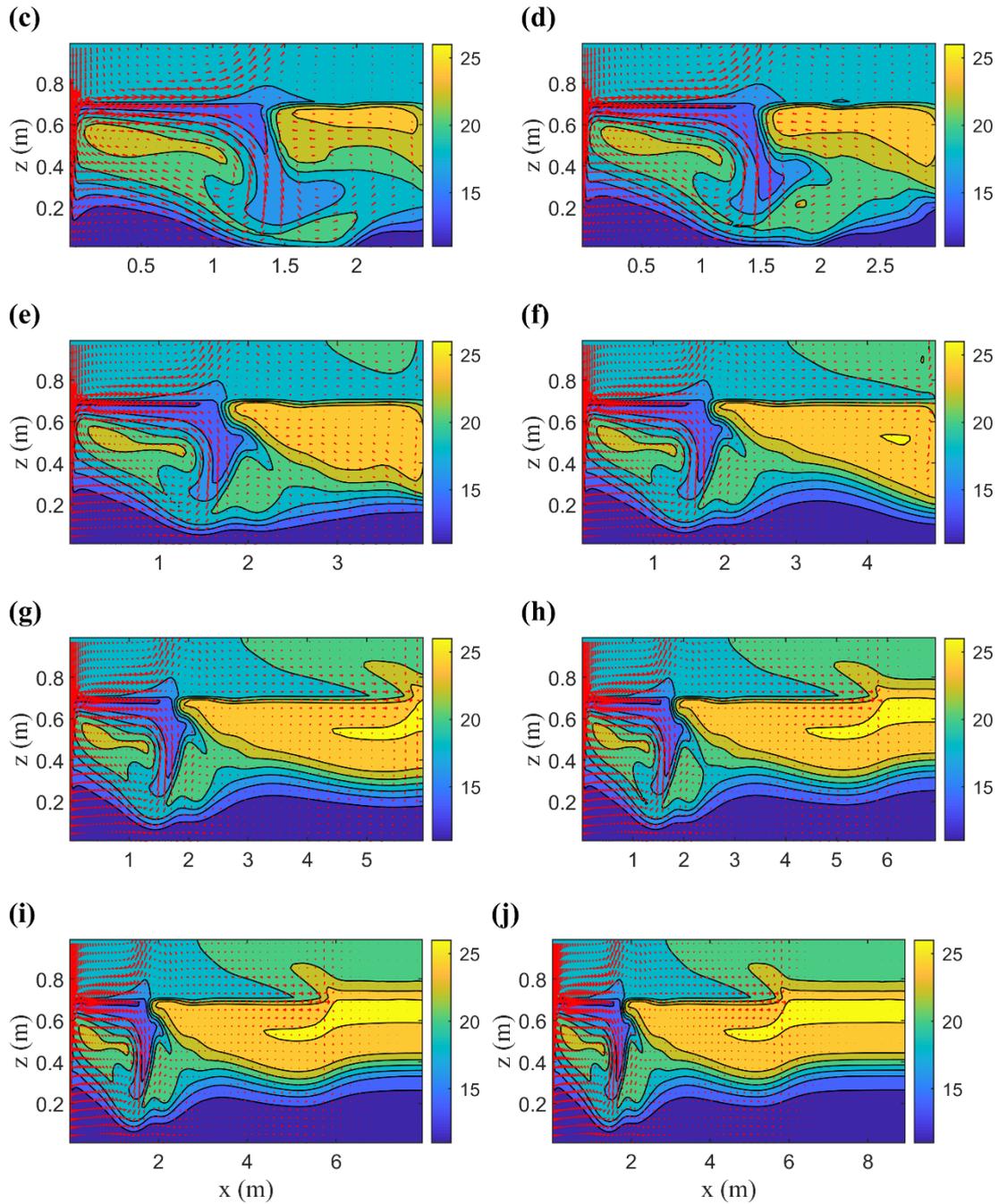

Fig. 20 The water temperature structures of different water areas including $x=$ (a) 1.0, (b) 2.0, (c) 2.5, (d) 3.0, (e) 4.0, (f) 5.0, (g) 6.0, (h) 7.0, (i) 8.0 m and (j) 9.0 m, when the intrusion is formed.

It can be seen from Fig. 21 that when the distance from aeration location is less than 1.0 m, the time needed for temperature stabilization has no significant change with the change of water area, and does not exceed 600 s. When the distance from aeration location is more than 1.2 m, the time needed increases and then decreases with the extension of water area, and generally is greater than the time needed within the distance of 1.0 m. As a result, it can be concluded that the influence area of single bubble plumes in this aeration condition is about 1.0 m, and the time needed for stabilization of water within this range will not have a significant difference with water area changing as long as the intrusion is formed away from the boundary. Furthermore, it is found that with the inner vortex getting smaller, its final range is 0-1.0 m, approximately $1.4R_s$, which can be considered as the influence area of the bubble plume in this setup.

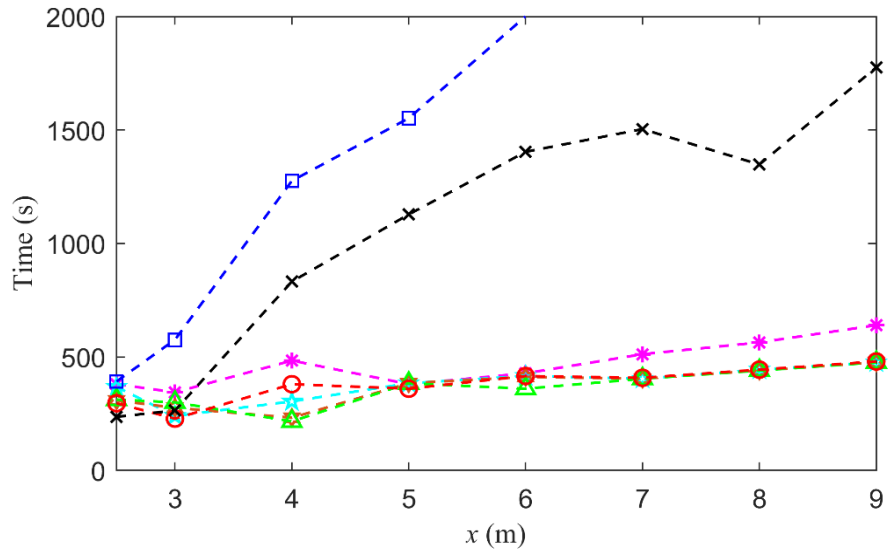

Fig. 21 Times needed for temperature stabilization of different sections which are 0.2 (dash-asterisk line), 0.4 (dash-pentagram line), 0.6 (dash-hexagram line), 0.8 (dash-triangle line), 1.0 (dash-circle line), 1.2 (dash-cross line) and 1.4 m (dash-square line) away from the aeration location, when $x \geq 2.5$ m.

## 5 Conclusion

In this study, a 3D two-phase mixture model combined with the energy model, NEWTANK, is developed to simulate the destratification in reservoirs induced by bubble plum. The model consists of Reynolds-Averaged Navier-Stokes Equations to investigate hydrodynamics characteristics, an advection-diffusion equation of air concentration to track the distribution, energy equations to solve the temperature field, and a $k$-$\varepsilon$ model to reflect the effects of aeration and water temperature on turbulence. The numerical model has been validated by existing experimental data and can be utilized as a useful tool in predicting the dynamics of bubble plume in reservoirs.

In addition, the effects of aeration rate, aeration location, and the water area on the destratification are investigated by utilizing the validated model with the following conclusions:

(1) When the aeration rate is low, bubble plumes are insufficient for effectively mixing the stratified water. It is generally believed that increasing the aeration rate reduces the time required to achieve well-mixed water. However, if the aeration rate is excessively high, it may actually prolong the time needed to reach stability. Therefore, a non-dimensional number $N_t$ is introduced, and the optimal aeration rate can be determined when the value of $N_t$ approaches 1.0.

(2) To ensure effective mixing, the aeration location should be positioned below the thermocline and as close to the bottom as possible.

(3) The water area can affect the destratification mechanism induced by bubble plumes. When water area is very small, causing the intrusion forms at or very close to the water boundary, there is only one vortex developed and the influence area of the bubble plumes is the whole water area. In contrast, when water area is relatively large to the extent that the intrusion forms away from the water boundary, two opposite vortexes are generated, with the influence area limited to the range where the inner vortex can develop, about $1.4R_s$.

However, there are still some limitations in this paper and the future work for improvement is outlined as follow. The equation for air volume fraction could be improved to consider the dissolution of air in water, the reaction among different components of air and the multiple bubble sizes. Additionally, factors such as evaporation, solar radiation and wind could be incorporated into the source terms of the energy model. The impact of multiple aeration devices should be studied, as a single aeration device may not be sufficient to fully mix stratified water in large lakes or reservoirs. It is aimed to get a standard method to assist the designer of bubble plume system in determining the optimal aeration condition, including the number of aeration device, aeration rate, the arrangement of aeration position and the time needed for temperature stabilization.

## Acknowledgments

This work was supported by the National Natural Science Foundation of China (Grant No. 52379077), and Open Fund of State Key Laboratory of Hydraulics and Mountain River Engineering (No. SKHL2208). The authors would like to express gratitude to Prof. Pengzhi Lin from Sichuan University in China for his valuable suggestions on this work.

# Disclosure statement

No potential conflict of interest was reported by the authors.

# Data availability statement

The data that support the findings of this study are available from the corresponding author upon reasonable request.